\documentclass[8.5pt,twoside,onecolumn]{article} 
\usepackage[super,sort&compress,comma]{natbib} 
\usepackage[version=3]{mhchem}

\usepackage{times,mathptmx} 
\usepackage{times}

\usepackage{sectsty} 
\usepackage{balance} 
\usepackage{graphicx} 

\usepackage{lastpage} 
\usepackage[format=plain,justification=raggedright,singlelinecheck=false, font=small,labelfont=bf,labelsep=space]{caption} 
\usepackage{fancyhdr} 
\title{O$_2$ adsorption trends on small supported PtNi clusters}

\author{Lauro Oliver Paz-Borb\'{o}n$^{1,2,\*}$ and Francesca Baletto$^1$}

\begin{document}

\makeatletter 
\def
\subsubsection{\@startsection{subsubsection}{3}{10pt}{-1.25ex plus -1ex minus -.1ex}{0ex plus 0ex}{\normalsize\bf}} 

\renewcommand{\headrulewidth}{1pt} 
\renewcommand{\footrulewidth}{1pt} \setlength{\arrayrulewidth}{1pt} \setlength{\columnsep}{6.5mm} \setlength\bibsep{1pt}

\maketitle

\begin{abstract}
We present a systematic analysis of molecular oxygen (O$_2$) adsorption trends on bimetallic Pt-Ni clusters and their monometallic counterparts supported on MgO(100), by means of periodic DFT calculations for sizes between 25 up to 58 atoms. O$_2$ adsorption was studied on a variety of inequivalent sites for different structural motifs, such as truncated octahedral (TO), cuboctahedral (CO), icosahedral (Ih) and decahedral (Dh) geometries. We found that O$_2$ prefers to bind on top of two metal atoms, parallel to the cluster, with an average chemisorption energy of 1.09 eV (Pt-Ni), 1.07 eV (Pt) and 2.09 eV (Ni), respectively. The largest adsorption energy values are found to be along the edges between two neighbouring (111)/(111) and (111)/(100) facets; while FCC and HCP sites located on the (111) facets may show a chemisorption value lower 0.3 eV where often fast O$_2$ dissociation easily occurs. Our results show that, even though it is difficult to disentangle the geometrical and electronic effects on the oxygen molecule adsorption, there is a strong correlation between the calculated general coordination number (GCN) and the chemisorption map. Finally, the inclusion of dispersion corrections (DFT-D) leads to an overall increase on the calculated adsorption energy values but with a negligible alteration on the general O$_2$ adsorption trends. 
\end{abstract}

\footnote{Department of Physics, King's College London, WC2R2LS, London}
\footnote{Instituto de F\'{i}sica, Universidad Nacional Aut\'{o}noma de M\'{e}xico, Apartado Postal 20-364, M\'{e}xico DF 01000, M\'{e}xico. email: oliver$\textunderscore$paz@fisica.unam.mx }

\section{Introduction}

\begin{figure*}
	[h] 
	\begin{center}
		\scalebox{0.26}[0.26]{
		\includegraphics[bb= 0 0 1300 1300 angle=0]{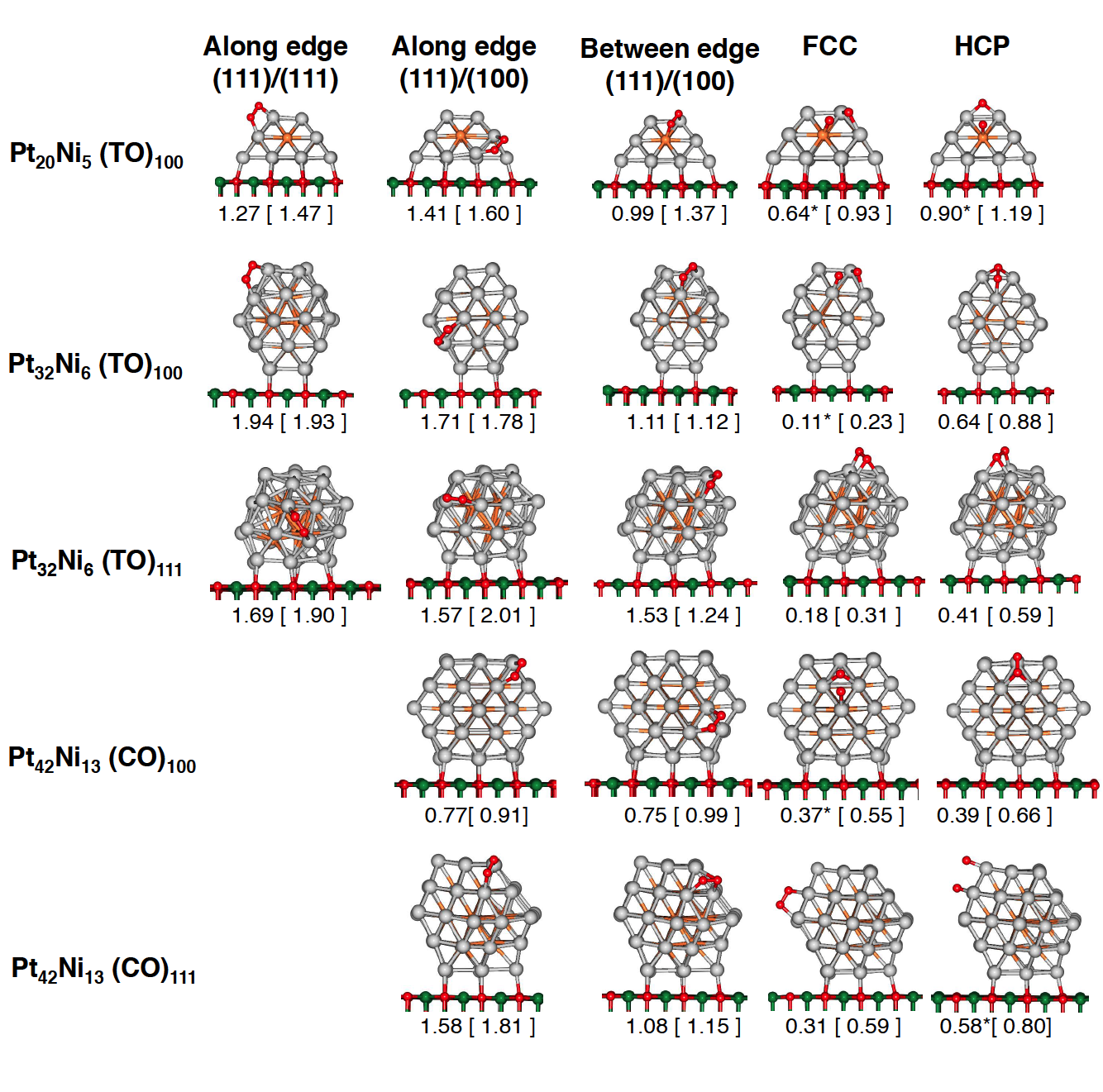}} \\
	\end{center}
	\caption{ Inequivalent O$_2$ chemisorption sites considered for both truncated octahedral (TO) and cuboctahedral (CO) Pt-Ni clusters, ranging from 25 to 55 atoms, supported on MgO(100). A distinction is made if the cluster is in contact with the substrate via one of its (111) or (100) facets. Calculated E$_{chem}$ values (PBE) are displayed in eV, where dispersion corrected (DFT-D) values are shown in brackets. An asterisk (*) is place on those configurations where O$_2$ was dissociated after DFT relaxation. Colour labelling: light grey (Pt), orange (Ni), green (Mg) and red (O). } \label{fig:sites1} 
\end{figure*}
\begin{figure*}
	[h] 
	\begin{center}
		\scalebox{0.26}[0.26]{
		\includegraphics[bb= 0 0 1200 900 angle=0]{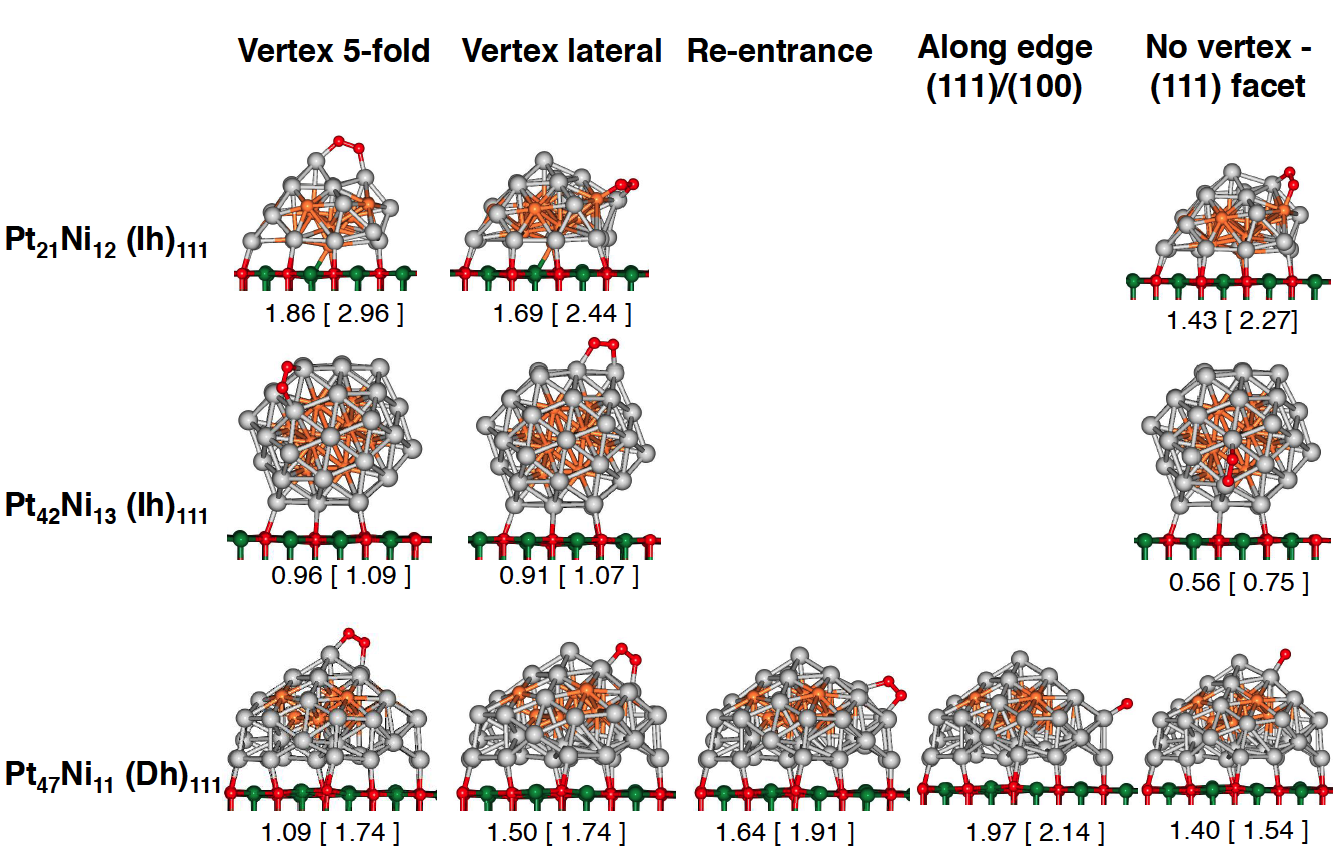}} \\
	\end{center}
	\caption{ Inequivalent O$_2$ adsorption sites on non-crystallographic icosahedral (Ih) and decahedral (Dh) Pt-Ni clusters, ranging from 33 to 58 atoms, supported on MgO(100). Calculated E$_{chem}$ values (PBE) are displayed in eV, where dispersion corrected (DFT-D) values are shown in brackets. Colour labelling: light grey (Pt), orange (Ni), green (Mg) and red (O). } \label{fig:sites2} 
\end{figure*}

The development of novel mobility technologies for cleaner vehicle emissions is essential to mitigate the current high levels of pollution seen on internal combustion engines. One foreseeable technology involves the electrochemical conversion of energy using polymer electrolyte membrane fuel cells (PEMFC)\cite{ShaoCHEMICALREVIEW2016, DebeNATURE2012}. Used in cars, fuel cells currently provide the necessary power needed to travel even long-distances. However, the costly use of platinum (Pt) as an electrocatalyst inside PEMFC has triggered an a vast active search for cheaper Pt-based alloys. One proposed solution is to combine it with other late transition metals such as nickel (Ni), cobalt (Co), chromium (Cr), copper (Cu) and iron (Fe)\cite{GreeleyNChem2009, StrasserNChem2010, StamenkovicNMAT2007}. The goal of the alloyed material is to improve the catalytic activity towards the oxygen reduction reactions (ORRs) offered by commercial Pt cathode catalysts without the scaling costs. In this respect, recent studies have shown that a five-fold decrease in the amount of Pt currently used in PEMFC stacks is needed, in terms of Pt costs and scarcity, in order to reach levels for mass-production and commercialisation in light-duty vehicles \cite{GasteigerAPPLIEDCatalysis2005}. Among these alloys, a promising one corresponds to Pt-Ni. Recent experimental studies performed by Stamenkovic \textit{et al.} shown an exceptional catalytic activity (10-fold) towards ORR of an extended Pt$_3$Ni(111) surface compared to a monometallic Pt(111) surface; and up to 90 times than the carbon supported Pt catalysts used in PEMFC \cite{StamenkovicSCIENCE2007}. This was attributed to an unusual shift in the Pt-Ni \textit{d}-band centre along with a peculiar atomic segregation pattern involving a Pt-rich outer layer. Furthermore, the maximisation of the large surface areas offered by nanoparticles (NPs) makes them ideal as catalysts for fuel cells applications. Recent experimental works have focused on the preparation of small Pt-Ni NPs ($\leq$ 5 nm) having well-controlled octahedral geometries (FCC-type atomic arrangements) along with extensive (111) facets aiming for similar catalytic properties as the extended surface\cite{StrasserSCIENCE2015, CarpenterJACS2012, CuiNATURE2013, ChoiACSNANO2014, ChoiNanoLetters2013, CuiNanoLett2012, HuangENERGY2014, ZhangMATERIALSCHEM2014, ZhangJACS2014, BaoJPCC2013}. 

From a theoretical view, computational Monte Carlo (MC) simulations have previously addressed the structural stability of cuboctahedral Pt-Ni NPs\cite{WangMONTECARLO2005}. Using many-body potential for particles ranging from 2.5 to 5 nm. MC simulations at 600K have shown that Pt-Ni tend to form \textit{surface-sandwich} structures, with a segregation pattern in which Pt atoms are enriched the outermost and third shells while the Ni atoms are enriched in the second shell. These results suggest an economical catalyst design with Pt atoms located at the outermost shells with Ni occupying core positions. Similar mixed segregation patterns were obtained, though for small Pt-Ni clusters of less than 20 atoms, using a genetic algorithm global optimization approach via density functional theory (DFT) calculations \cite{GarzonPTNI2009}, while more recent work, we have shown that sub-nanometre gas-phase Pt-Ni clusters up to 55 atoms can bind $O_2$ too strongly (above 1 eV) due to substantial geometrical reconstruction of the metal-metal bridge underneath the molecule, making those clusters not ideal for catalysing the sluggish ORR compared to a Pt(111) surface\cite{ConoPCCP2011}. More recent DFT calculations performed by Fortunelli and co-workers have reported a fully-dealloyed Pt$_3$Ni$_7$ particule ($\sim$ 8 nm.) surfaces exhibiting triangulated surface arrangements, as a regular Pt(111), while reducing the rate-determining ORR step significantly\cite{FortunelliCHEMSCIE2015}.

In this work we performed extensive DFT periodic calculations to quantify O$_2$ adsorption on a variety of Pt-Ni clusters supported on MgO(100) as well as on their monometallic (Pt, Ni) counterparts. Using the calculated adsorption energies (E$_{chem}$) we are able to construct a O$_2$ chemisorption map over supported Pt-Ni, Pt and Ni clusters, relevant to fuel cells and heterogeneous catalysis. Furthermore, the interfacial low-index surface displayed at the cluster/oxide interface - where the cluster can expose both (100) and (111) facets in contact with the oxide interface - was also analysed. For the bimetallic case, attention is placed on Ni doping at clusters core sites (\textit{i.e.} core-shell clusters), though in some cases Ni atoms are located inevitably located at surface sites due to cluster size (TO$_{25}$) or geometrical arrangements (Dh$_{58}$). Overall, we identify the (111)/(111) and (111)/(100) cluster edges as those sites were calculated E$_{chem}$ values are the strongest (above 1 eV) which tend to have a low GCN value ($< $6); while FCC and HCP sites located at (111) facets have the weakest E$_{chem}$ values (less than 0.64 eV) involving GCN $>$ 8 . As GCN can distinguish symmetrically equivalent sites, it allow us to rationalise calculated E$_{chem}$ values among different clusters since weaker interactions occur at sites where GCN numbers are higher. It has been recently shown that GCN further establishes a link between geometry, adsorption and activity\cite{CallePHYSCHHEMLETT2014, CalleANGEWANDTE014, CalleNATURECHEM2015, CalleSCIENCE2015}.

\section{Methodology}

We considered 4 different structural motifs ranging from FCC-type such as truncated octahedra (TO), cuboctahedral (CO), and non-crystallographic arrangements such as icosahedral (Ih) and decahedral (Dh); with sizes ranging from 25 up to 58 atoms. These structures are constructed from highly symmetric models: 38(TO)$_{\alpha}$, 55(CO)$_{\alpha}$, 55(Ih)$_{\alpha}$ and 75(Dh)$_{\alpha}$. Half-cuts were performed on these structures to obtain a new set of clusters with exotic cluster/oxide interfaces: 25(TO)$_{\alpha}$, 33(Ih)$_{\alpha}$ and a 58(Dh)$_{\alpha}$. Here, the subscript ${\alpha}$ stands for the Miller indexes of the cluster-oxide interface initially exposed. In the following, both (100) or a (111) facets have been chosen. Pt-Ni clusters display a deliberate Pt$_{shell}$-Ni$_{core}$ chemical arrangement. Depending on size and structural, some Ni atoms are find to occupy surface sites, for example Pt$_{20}$Ni$_{5}$(TO)$_{100}$, Pt$_{21}$Ni$_{12}$(Ih)$_{111}$ as well as Pt$_{47}$Ni$_{11}$(Dh)$_{111}$ (see Figs. \ref{fig:sites1} and \ref{fig:sites2}). After construction in the gas-phase, the metal clusters are subsequently placed over the MgO(100) substrate maximising the number of metal-oxygen (M-O) bonds. An initial height of $\sim$ 2 \AA\ prior DFT relaxations is employed between the cluster and the substrate (optimal Pt and Ni atoms DFT heights adsorbed on O-sites surface sites are 1.985 \AA\ and 1.796 \AA, respectively). The substrate thus acts as a geometrical constraint affecting the features of each metallic clusters. As part of the cluster/oxide interfacial geometry characterisation, we have counted the number of metal atoms at the interface. Furthermore, as a measure of the metal interface \textit{roughness}, we have calculated the standard deviation of the corresponding interfacial atomic layer height, information which can be found in the Supplementary Material along with other calculated quantities. We must stress that the considered supported Pt, Ni and Pt-Ni clusters, though locally DFT-relaxed, they do not correspond to global minima in the potential energy surface (PES). However, they act as geometrical scenarios to understand their O$_2$ adsorption properties. 

The plane-wave implementation of Density Functional Theory (DFT) within the Quantum Espresso (QE) code \cite{QUANTUM_REF} with the PBE exchange-correlation (xc) functional\cite{PBEfunctional1996}. A combination of Rappe-Rabe-Kaxiras-Joannopoulos \cite{RappePSEUDOPOT1991} and Vanderbilt\cite{VanderbiltPSEUDOPOT2014} ultrasoft pseudopotentials - including non-linear core corrections - are use to describe Pt(5d$^9$6s$^1$), Ni(3d$^9$4s$^1$), Mg(2p$^6$3s$^1$3p$^{0.75}$) and O(2s$^2$2p$^4$) atoms. A kinetic energy cutoff for wave-functions and charge density of 45 and 360 Ry, respectively is used. To improve convergence, a Marzari-Vanderbilt smearing value of 0.001 Ry is used. The pristine MgO(100) surface is modelled using a 6$\times$6 three-layer slab our calculated PBE bulk lattice constant (4.238 \AA). Although this is slightly larger than the experimental value of 4.21\AA, is whoever, in close agreement with calculated PBE values (4.26 and 4.30 \AA) and hybrid (PBE0) functionals (4.21 \AA)\cite{Broqvist2004, Paier2006MgO}. The MgO(100) substrate has been considered on the basis that it is a often used in experiments, it is a well-characterized and rigid substrate strongly interacting with a metallic cluster and it can be accurately treated with standard ab-initio methods. Its checkerboard (100) surface allow us to analyse any effects between flat and corrugated oxide-cluster interfaces on the calculated E$_{chem}$ values. The size of the supercell (17.9803 \AA $\times$ 17.9803 \AA $\times$ 29.2380 \AA) provides a sufficiently large vacuum region, necessary to avoid any spurious interactions between neighbouring periodic images. All DFT calculations have been performed spin-polarised. The Brillouin zone was sampled at the Gamma point only. Empirical dispersion corrections are included via single-point calculations on PBE relaxed configurations using the DFT+D functional, which includes a pairwise addition C$_6$R$^6$ correction term\cite{GrimmeCOMPCHEM2004}. O$_2$ adsorption (chemisorption) energies ($E_{chem}$) are calculated as total energy differences between the interacting configuration (E$_{O_2 + cluster/MgO}$), the lowest-energy supported bare cluster (E$_{cluster/MgO}$), and the O$_2$ molecule in the gas-phase: 
\begin{equation}
	E_{chem} = - (E_{O_2 + cluster/MgO} - E_{cluster/MgO} - E_{O_{2, gas phase}}) \label{eq1} 
\end{equation}

where positive E$_{chem}$ values indicate a exothermic O$_2$ adsorption. We also quantify the Pt-Ni, Pt and Ni clusters \textit{core} and \textit{shell} strain as the percentage difference between the average nearest-neighbour (NN) distance for those atoms occupying core (d$_{NN}^{c}$) and shell (d$_{NN}^{c}$) positions with respect to PBE calculated Pt (2.812 \AA) and Ni (2.487 \AA) lattice constants, while for Pt-Ni we considered the Ni-Pt$_3$ phase with an experimental value of 2.718 \AA, see ref.\cite{EngelkePHD2010} (d$_{NN}^{PBE}$). Negative values in Eq. \ref{eq2} indicate a compressive strain felt by the cluster: 
\begin{equation}
	\begin{array}{cc}
		s_{core} = {\frac{d_{NN}^{c} - d_{NN}^{PBE}}{d_{NN}^{PBE}}} \times 100 \\
		\\
		s_{shell} = {\frac{d_{NN}^{s} - d_{NN}^{PBE}}{d_{NN}^{PBE}}} \times 100 \\
	\end{array}
	\label{eq2} 
\end{equation}

We employ another quantity to geometrically characterise the adsorption site, the generalized coordination number (GCN). Introduced by Sautet and co-workers for monometallic Pt systems the GCN is a new descriptor as robust as the \textit{d}-band center, which establishes a direct link between geometry, adsorption map and activity\cite{CallePHYSCHHEMLETT2014, CalleANGEWANDTE014, CalleNATURECHEM2015, CalleSCIENCE2015}. It to help us differentiate each adsorption site despite size, shape and composition of the cluster systems in order to highlight adsorption trends across a wide variety of structures. The GCN can be defined as the coordination number at the adsorption site weighted by the overall coordination of the neighbouring atoms: 
\begin{equation}
	GCN(i) = \sum_{j=1}^{n_i} CN(j) / CN_{max} \label{eq3} 
\end{equation}

where the sum runs over all neighbouring atoms \textit{n$_i$} of the atom \textit{i} and each neighbour is weighted according to its coordination number divided by the maximum coordination number of the adsorption site. We keep \textit{CN$_{max}$ } = 18 as the O$_2$ molecule is adsorbed at a bridge site (on top of 2 Pt$_{shell}$ atoms), as shown recently for larger supported Pt-Ni clusters\cite{AsaraACSCATALYSIS2016}.\\
\begin{figure*}
	[h] 
	\begin{center}
		\scalebox{0.2}[0.2]{
		\includegraphics[bb= 0 0 1600 1200 angle=0]{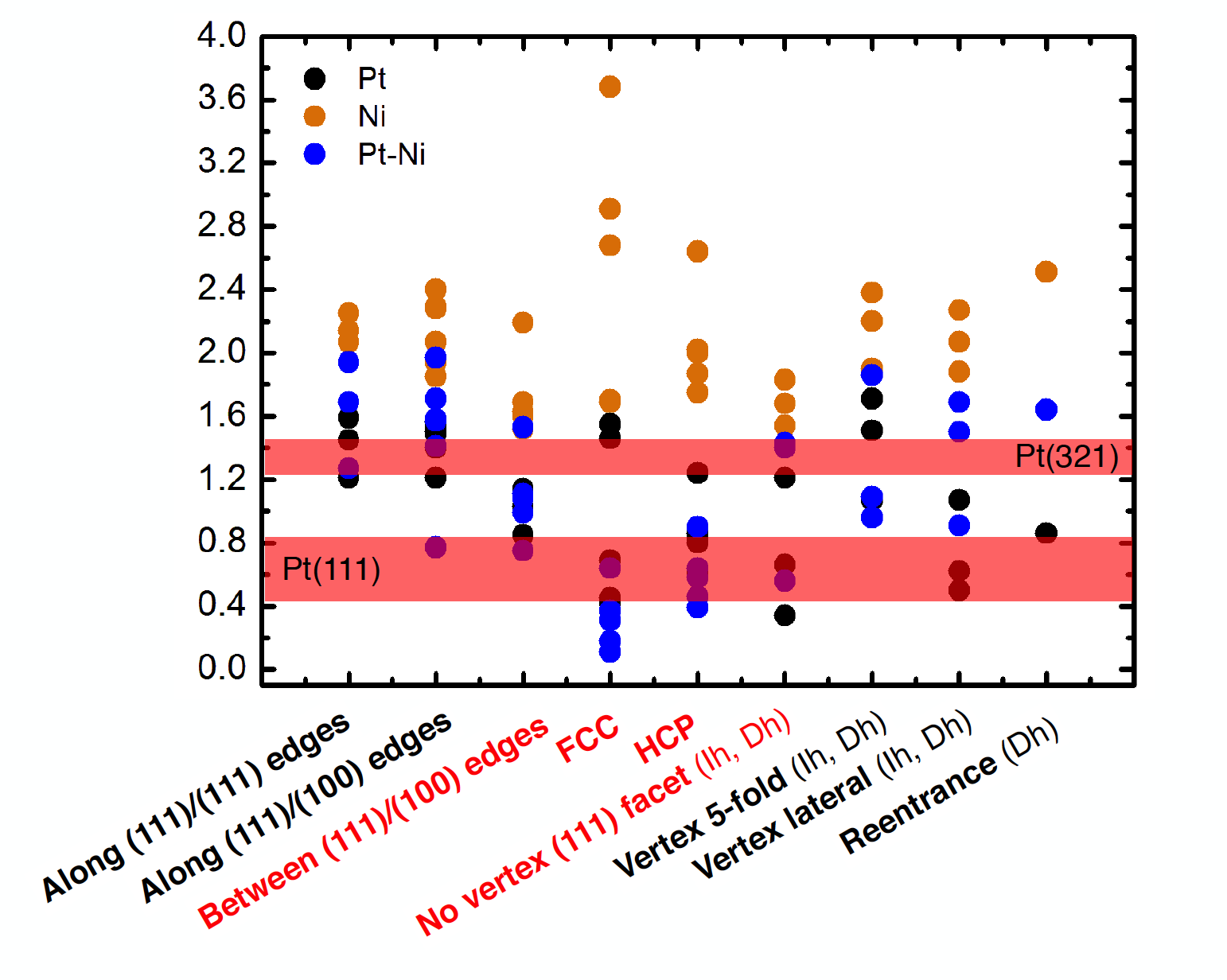}} 
	\end{center}
	\caption{ Calculated O$_2$ adsorption (E$_{chem}$) values at the PBE level for Pt, Ni and Pt-Ni clusters ($\leq$58 atoms) supported on MgO(100), including all the inequivalent adsorption sites shown in Figs \ref{fig:sites1} and \ref{fig:sites2}. PBE reference values for the O$_2$ adsorption energy range for the flat Pt(111) are highlighted by the transparent red area (0.46 up to 0.86 eV)\cite{GlandSURFSCIE1980, SteiningerSURFSCIE1982A, SautetPRB1999, EichlerPRL1997, JenningsNANOSCALE2014, McEwenPCCP2012}, as well as those for the stepped Pt(321) surface (1.23 to 1.56 eV)\cite{BrayLANGMUIR2011}. Colour labelling: black (Pt), orange (Ni), blue (Pt-Ni). } \label{fig:Echems} 
\end{figure*}

\section{Results and Discussion}

\vspace{0.0cm}
\begin{figure*}
	[htb] \centering 
	\begin{tabular}
		{@{}cc@{}} \scalebox{0.10 }[0.10 ]{
		\includegraphics[bb= 0 0 1500 1100 angle=0]{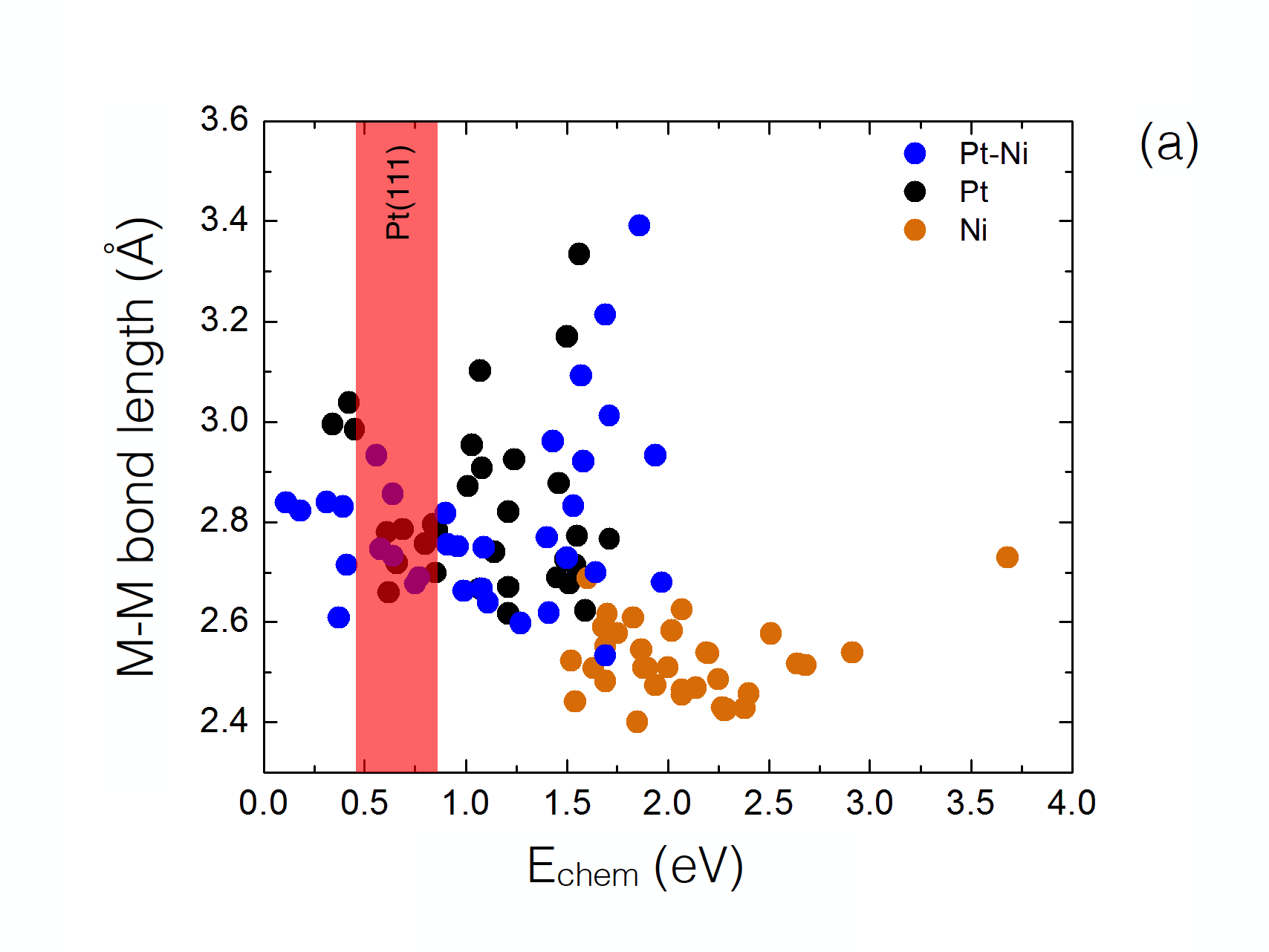}} & \scalebox{0.10 }[0.10 ]{
		\includegraphics[bb= 0 0 1400 1100 angle=0]{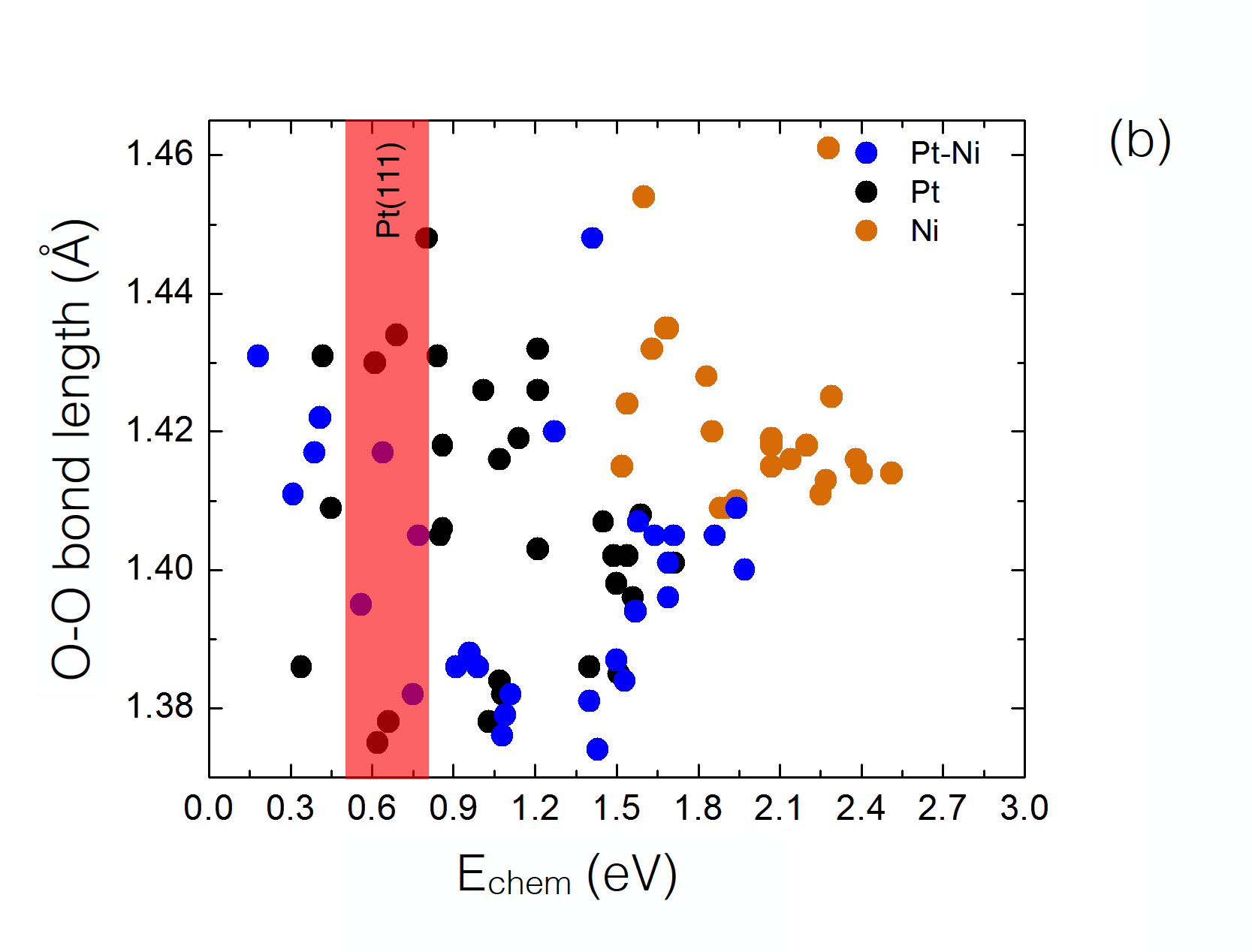}} \\
		\scalebox{0.10 }[0.10 ]{
		\includegraphics[bb= 0 0 1500 1100 angle=0]{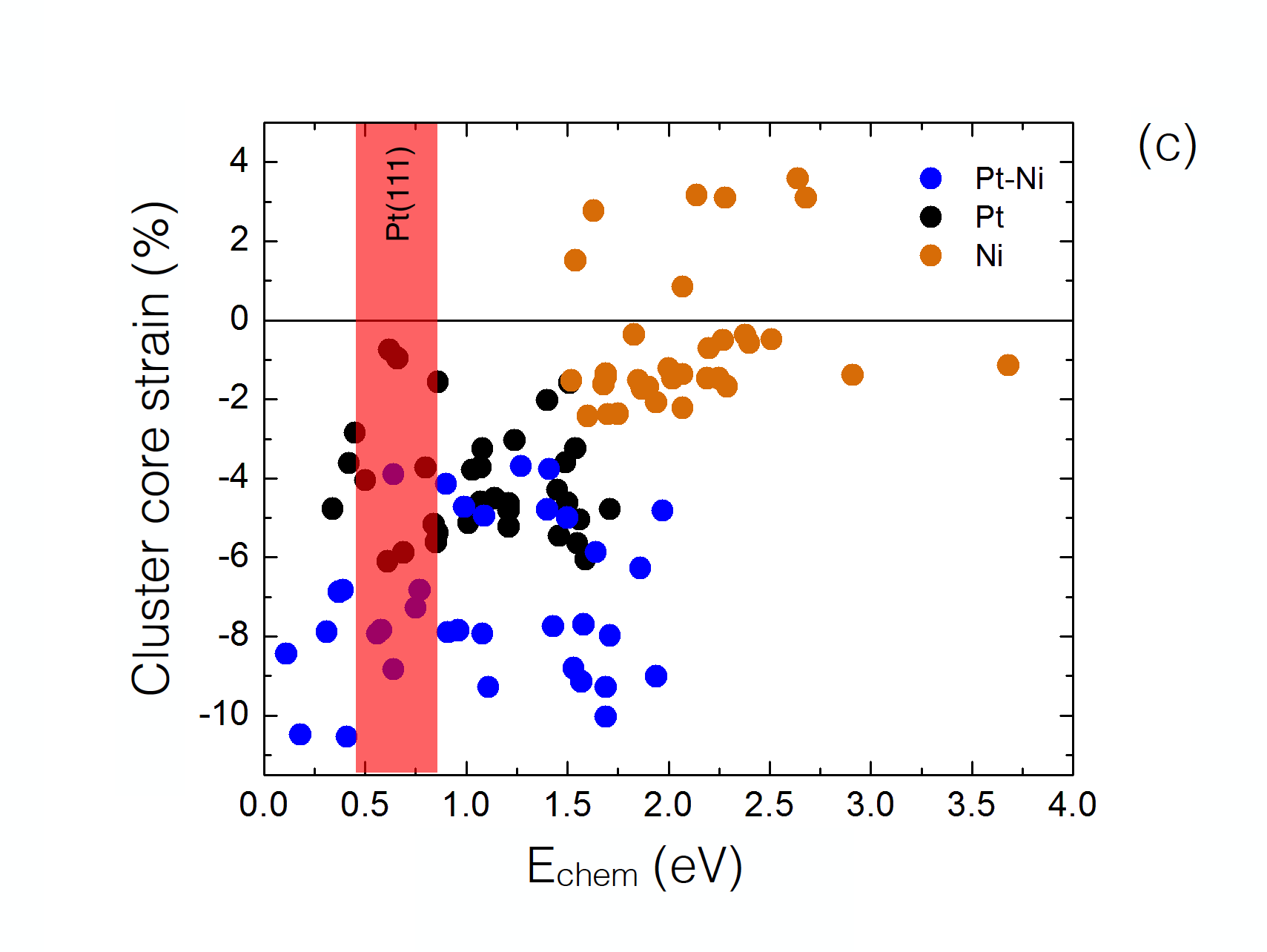}} & \scalebox{0.10 }[0.10 ]{
		\includegraphics[bb= 0 0 1500 1100 angle=0]{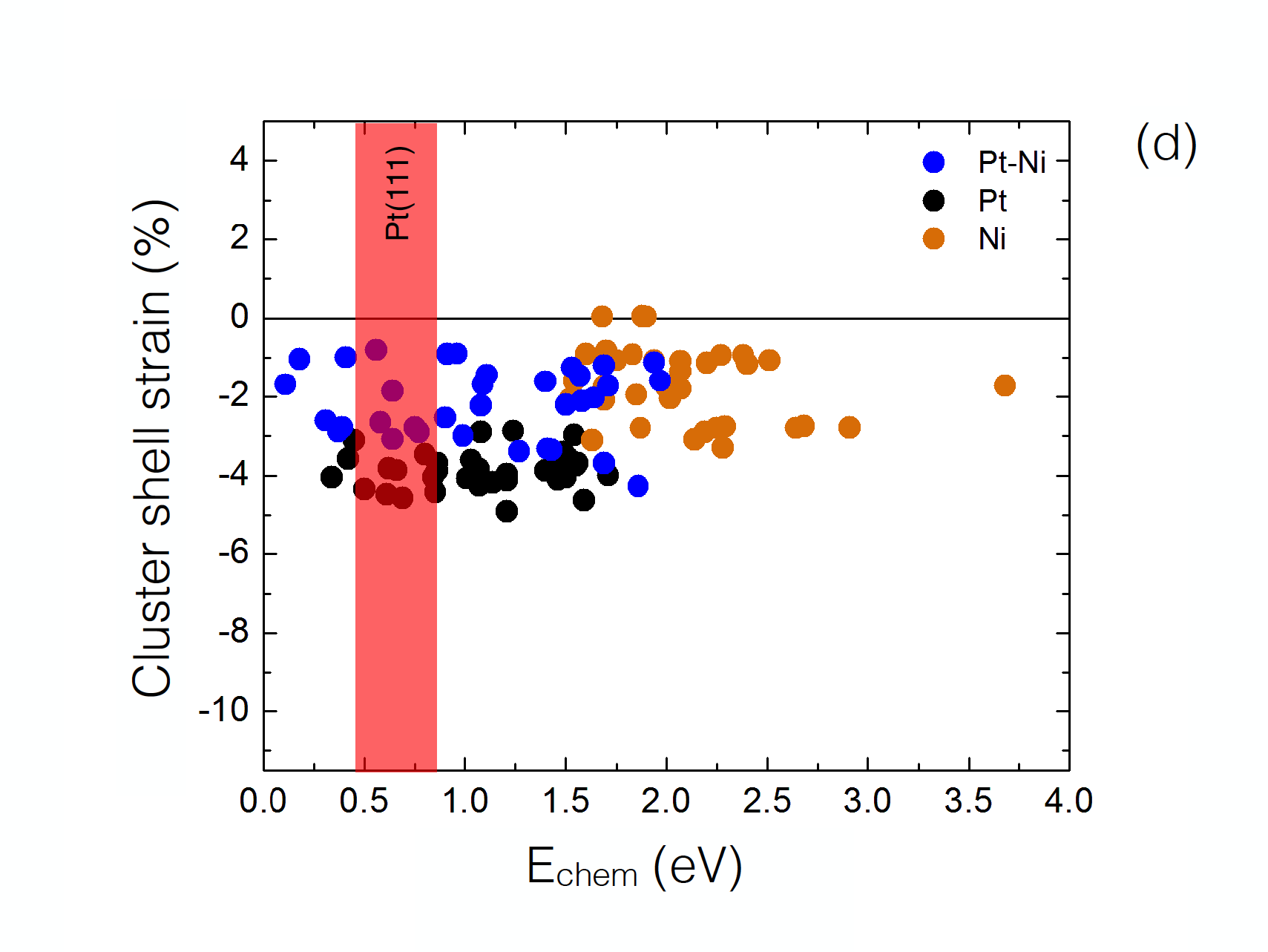}} \\
		\scalebox{0.10 }[0.10 ]{
		\includegraphics[bb= 0 0 1500 1100 angle=0]{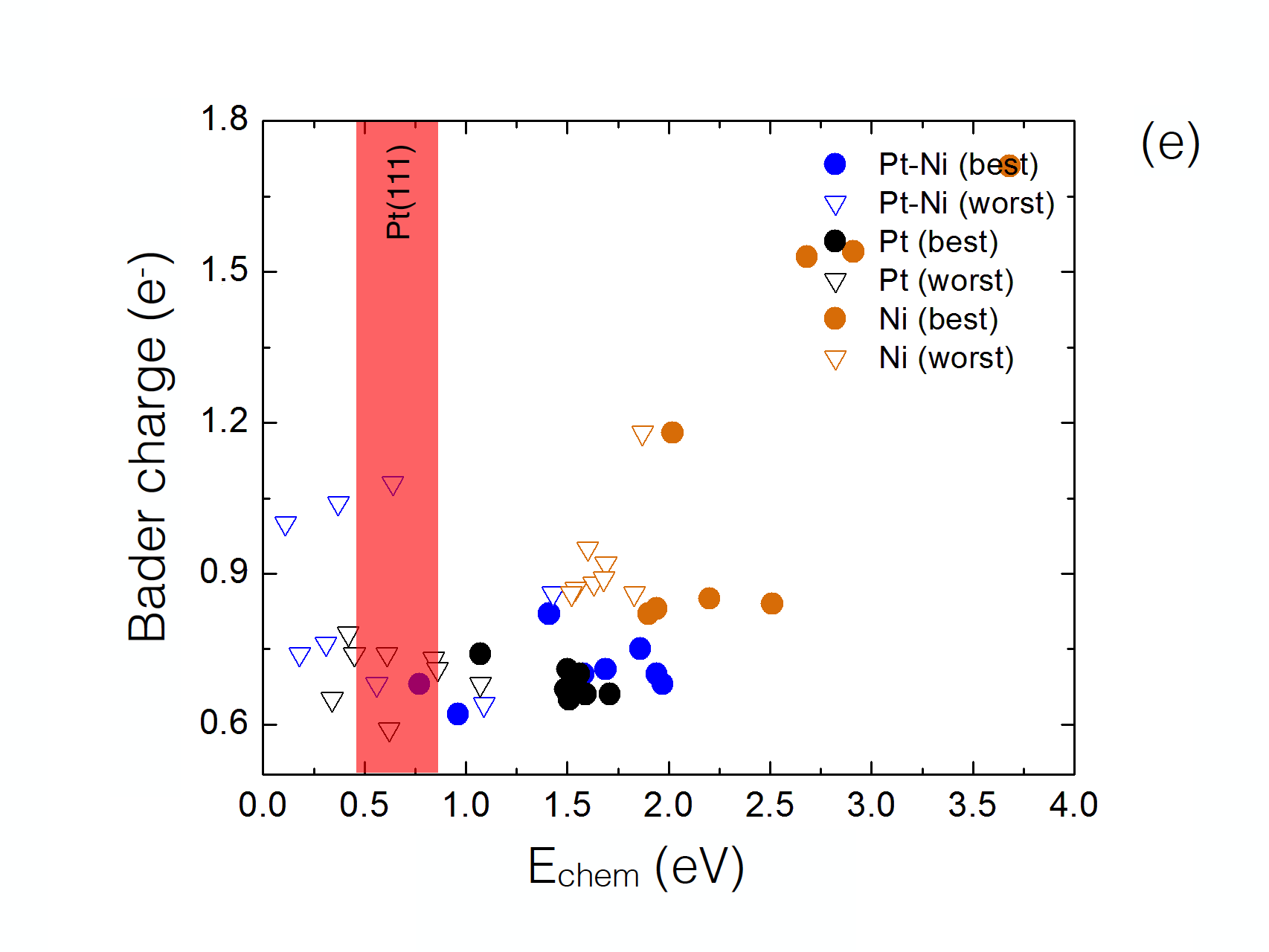}} & \scalebox{0.10 }[0.10 ]{
		\includegraphics[bb= 0 0 1500 1100 angle=0]{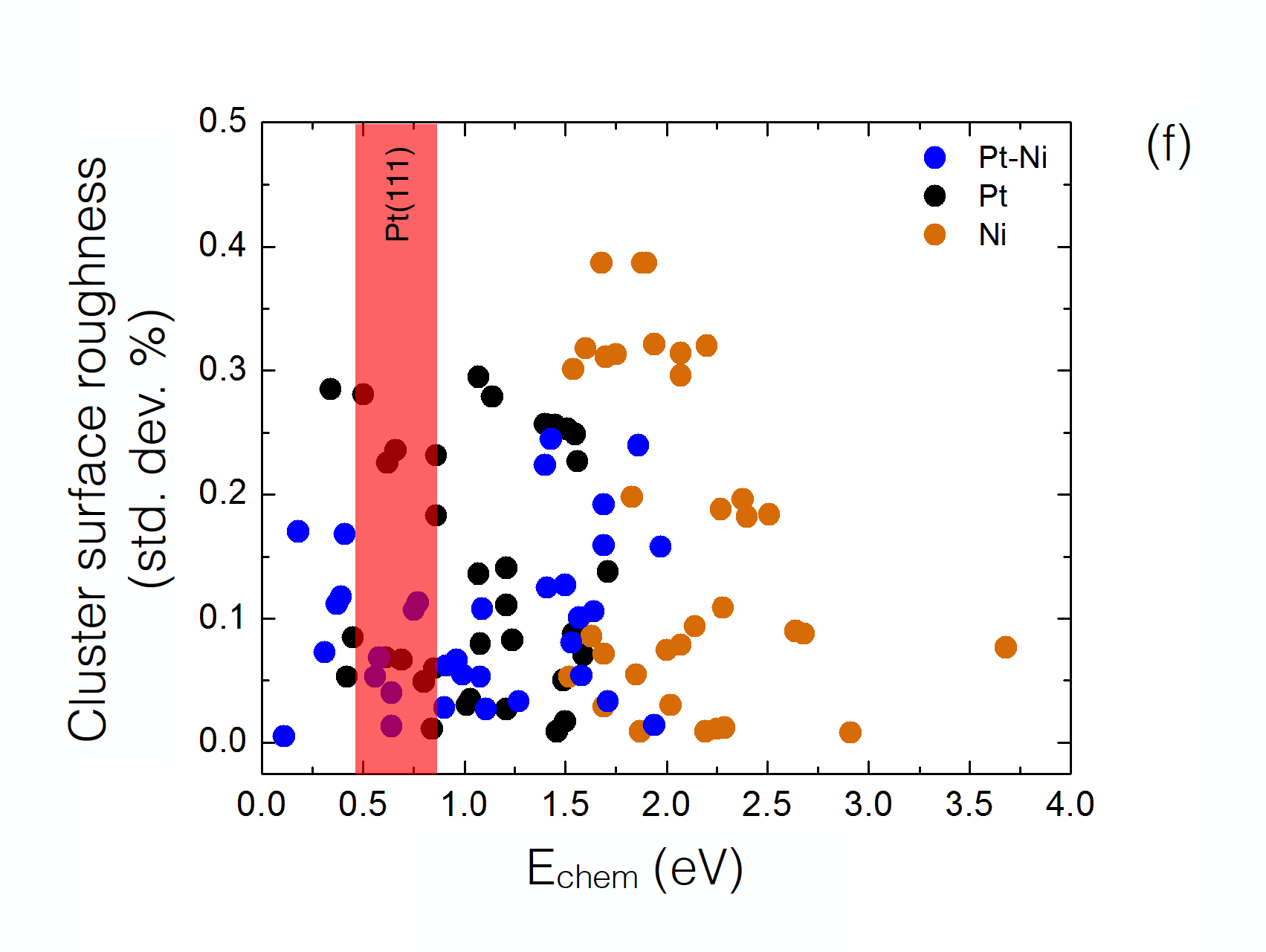}} \\
	\end{tabular}
	\caption{Structural analysis of all O$_2$ adsorption sites on Pt-Ni as well as on the monometallic Pt, Ni supported clusters, displaying calculated PBE E$_{chem}$ values as function of: (\textit{a}) the metal-metal (M-M) bond length at the adsorption site, (\textit{b}) the O$_2$ molecule bond length, (\textit{c} and \textit{d}) cluster \textit{core} and \textit{shell} strain, (\textit{e}) calculated cluster Bader charge transfer (both \textit{best} and \textit{worst} adsorption sites) and (\textit{f}) the cluster surface \textit{roughness}. Experimental and PBE reference values for the O$_2$ adsorption energy range for the flat Pt(111) are highlighted by the transparent red area (0.46 up to 0.86 eV)\cite{GlandSURFSCIE1980, SteiningerSURFSCIE1982A, SautetPRB1999, EichlerPRL1997, JenningsNANOSCALE2014, McEwenPCCP2012}.} \label{fig:Data} 
\end{figure*}

Due to the wide variety of Pt-Ni geometries and sizes we were able to identify several inequivalent sites for O$_2$ adsorption. These O$_2$ adsorption sites have been grouped into different \textit{families} as shown in both Figs \ref{fig:sites1} and \ref{fig:sites2}. Prior to DFT relaxations, the O$_2$ molecule was always placed horizontally towards the cluster surface, (\textit{bridging} between two metal atoms), as recent studies theoretical studies show that this is the preferred adsorption configuration \cite{TaoJPCB2001, JenningsNANOSCALE2014, ShaoNANOLETTERS2011}. Our analysis thus focuses mostly on the differences between the strongest and weakest O$_2$ adsorption sites found for supported Pt-Ni clusters and use the bare Pt and Ni cases as references.

\subsection{Overall adsorption trends}

\begin{figure}
	\begin{center}
		\scalebox{0.35}[0.35]{
		\includegraphics[bb= 0 0 800 1200 angle=0]{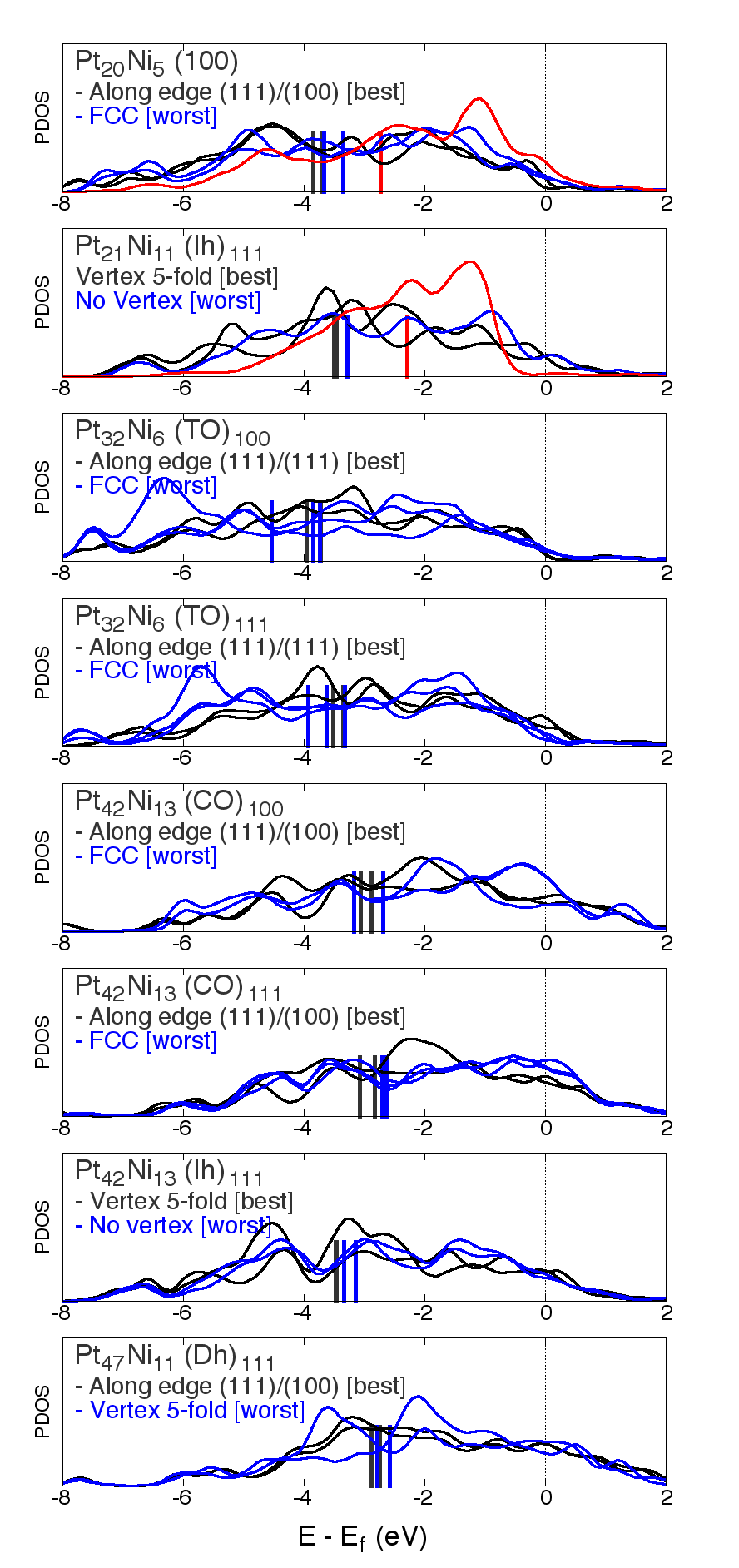}} \\
	\end{center}
	\caption{ Calculated PBE projected density of states (PDOS) for supported Pt-Ni clusters (25 to 58 atoms), where the position of the \textit{d}-center is shown for the \textit{best}(stronger) and \textit{worst} (weakest) O$_2$ adsorption sites, in black and blue lines, respectively. O$_2$ adsorption involving Ni atoms are shown in red lines.} \label{fig:dcenters} 
\end{figure}
\begin{figure}
	\begin{center}
		\scalebox{0.2}[0.2]{
		\includegraphics[bb= 0 0 1400 1200 angle=0]{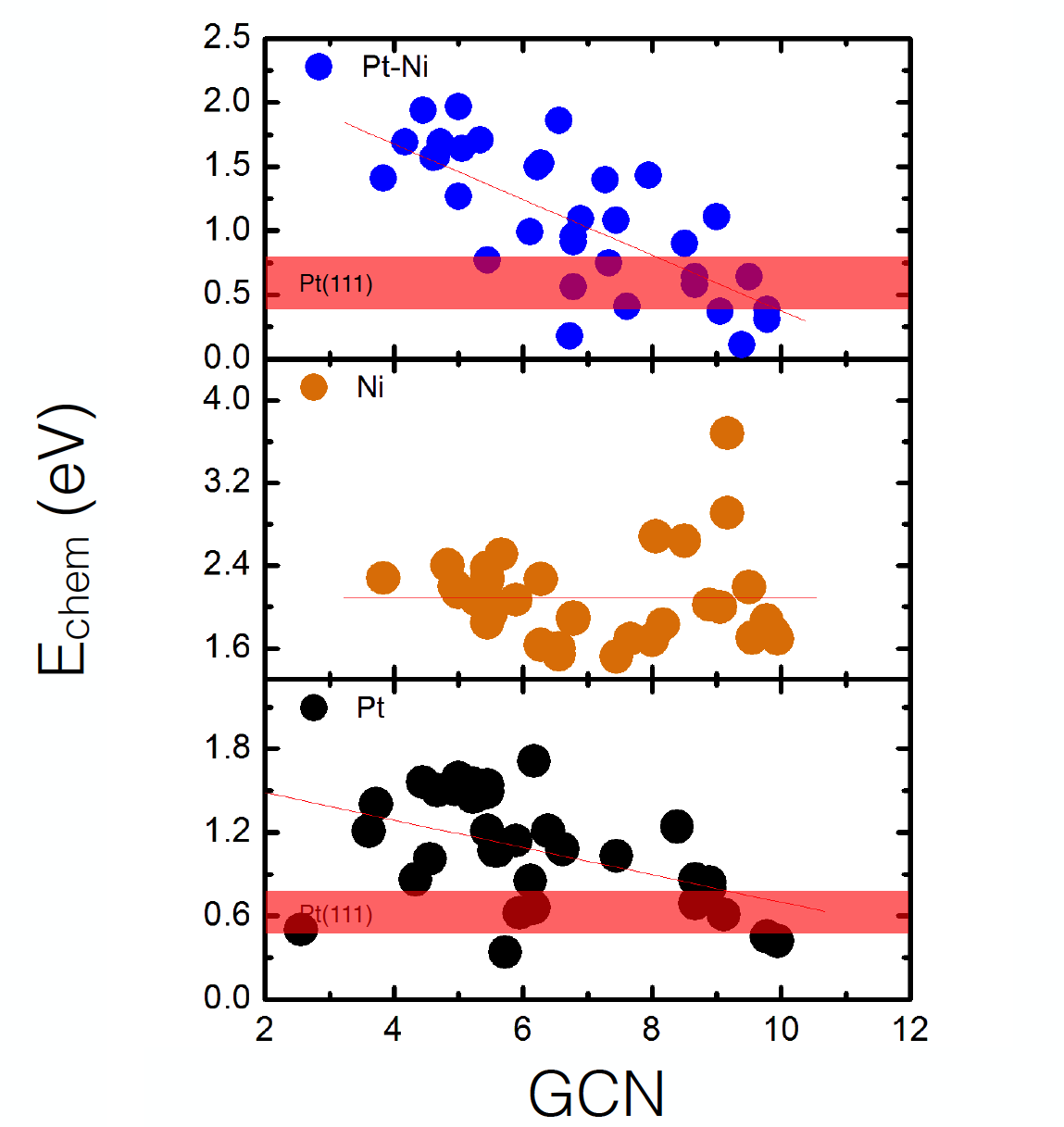}} \\
	\end{center}
	\caption{ Calculated GCN values for all inequivalent O$_2$ adsorption sites for all Pt-Ni, Pt and Ni supported clusters (sizes ranging from 25 up to 58 atoms), plotted as a function of calculated PBE E$_{chem}$ values.} \label{fig:GCN} 
\end{figure}

An O$_2$ chemisorption map over Pt, Ni and Pt-Ni supported clusters is shown in Fig.\ref{fig:Echems}. Here, calculated E$_{chem}$ values are plotted as a function of all inequivalent sites considered. Previously calculated PBE reference values for the O$_2$ adsorption on the flat Pt(111) are highlighted by the red area (0.46 up to 0.86 eV)\cite{SautetPRB1999, EichlerPRL1997, JenningsNANOSCALE2014, McEwenPCCP2012}, including experimental values (0.4 - 0.5 eV)\cite{GlandSURFSCIE1980, SteiningerSURFSCIE1982A}, as well as for the stepped Pt(321) surface (1.23 to 1.56 eV)\cite{BrayLANGMUIR2011}. From Fig.\ref{fig:Echems}, we identify 4 Pt-Ni adsorption sites which lead to E$_{chem}$ values resembling those of the Pt(111) surface: \textit{FCC}, \textit{HCP}, \textit{between edges}, (111)/(100) and \textit{no vertex} sites. In some cases, calculated E$_{chem}$ values are $<$ 0.5 eV, particularly at \textit{FCC}, \textit{HCP} sites, where in most of the DFT relaxations the O$_2$ molecule undergoes fast dissociation. We must stress we are only analysing the O$_2$ adsorption trends on a variety of inequivalent sites, in order to understand O$_2$ adsorption on a variety of Pt-Ni cluster structures. This represents the first step in the complex oxygen reduction reaction (ORR) which consisting of a series of protonation process of the dissociative products (O, OH, OOH and H$_2$O$_2$) to form a final H$_2$O product\cite{PEMFuelCells2008}. \\

The relevance of (111) facets on cluster structures as candidates for low-barrier O$_2$ dissociation has been recently discussed for truncated octahedral (TO) gas-phase clusters at atom sizes 38, 79 and 116 by Jennings \textit{et al}\cite{JenningsNANOSCALE2014, JenningsPCCP2014, JenningsJPCC2015}. According to their work, stronger $E_{chem}$ values are calculated for pure Pt$_{38}$ on both FCC and HCP sites (1.79 and 1.84 eV) compared to a core-shell platinum-titanium Pt$_{32}$Ti$_{6}$ bimetallic cluster (0.38 and 0.74 eV). However, they reported near barrier-free dissociation at the FCC/HCP sites for pure Pt clusters (0.00 and 0.04 eV) with slightly larger barriers for the bimetallic case (0.62 and 0.34 eV). The monometallic case behaviour was explained due to an easily distorted Pt(111) facet - namely the central Pt atom - facilitating O$_2$ dissociation; while an increase in the rigidity of this (111) facet was due to Ti-alloying at the core, for the bimetallic case. Similar trends were also reported for larger TO$_{79}$ particles\cite{JenningsNANOSCALE2014}. Overall, the rigidity of the (111) was reported to decrease for 3d metals from the 4-8 group (Ti to Fe), while similar distortions as of the pure Pt cluster were reported for groups 10-12 (Ni, Cu and Zn). \\

We monitored in detail the metal-metal (M-M) bond length at the O$_2$ adsorption site changes for the three different systems, Fig. \ref{fig:Data} (a). Upon O$_2$ adsorption, the Pt-Pt bond is stretched as a general trend due to a "softer" character of Pt, in order to accommodate the incoming O$_2$ molecule. This is noticeable at edge sites, sites such as \textit{along edges}, both (111)/(111) and (111)/(100); $E_{chem}$ values of 1.60 - 1.94 eV. Particularly at the TO$_{38}$ (111) facet, the central Pt atom involving in O$_2$ adsorption on Pt-Ni clusters is "pulled up". For Ni clusters, Ni-Ni bonds remain rather stable upon O$_2$ adsorption, clustered around 2.40 - 2.70 \AA, while it is the molecule O-O bond which is elongated - in some cases, stretched to the point of dissociation - as shown in Fig. \ref{fig:Data} (b). Having a Pt shell, the M-M bond length of bimetallic Pt-Ni clusters resemble those values calculated for monometallic Pt. From Fig.\ref{fig:Data} (a), we notice that the M-M bond length in Pt-Ni, Pt and Ni have a "L"-shaped distribution. A linear behaviour is observed where shortening the M-M bond stands for an E$_{chem}$ increment; when a critical compression is reached there is an abrupt change and small changes in the E$_{chem}$ are associated to very different M-M bond lengths. However, M-M bond elongation is not the sole cause of larger E$_{chem}$, but it is connected to both molecule O-O bond stretching and the adsorption site. Fig. \ref{fig:Data} (b), shows only the O-O bond length range for molecular adsorption. From this plot, it seems to be two regimes for Pt-Ni: one where the E$_{chem}$ values increase from 0.18 up to nearly 1.2 eV while the O-O bond decreases (1.43 \AA, down to 1.37 \AA), namely namely at FCC/HCP sites for TO/CO structures and the \textit{no-vertex} site of Ih clusters. A second regime, starting roughly at 1.3 eV up to 2.1 eV, is observed where the O-O bond length increases from 1.37 \AA up to 1.41 \AA, mostly \textit{along edges} and \textit{5-fold vertex} sites. However, this behaviour is more erratic for monometallic Pt clusters while Ni clusters involve O-O bond lengths with E$_{chem}$ above 1.5 eV.\\

Fig.\ref{fig:Data} (c) shows both \textit{core} and \textit{shell} calculated strain within the cluster, as it has been recently further shown a direct link between a nanoparticle (NP) reactivity and its compressive strain, particularly in bimetallic cases\cite{StrasserNChem2010, ShaoNANOLETTERS2011, HernandezNATURECHEM2014}. \textit{Core} strain for Pt and Pt-Ni bimetallic clusters is rather large, particularly for the bimetallic case where compressive values can reach reach values around -10$\%$ as this clusters are still small enough that have not yet developed bond lengths are those of the bulk. Ni clusters values show a modest compression ($<$ -2$\%$), except for the TO$_{25}$(100) structure, where 12 Ni atoms at the cluster/oxide interface need to match equal12 O surface sites, with average interface bond lengths of 2.52 \AA, compared to the average Ni-Ni cluster bond length of 2.45 \AA; thus enlarging the internal Ni-Ni core distances leading to positive core strain values. Similar trends are seen for the Ih$_{33}$(111), which due to its large and irregular cluster/oxide interface needs to stretch those Ni-Ni bonds at the interface and core positions. Yet, it is difficult to imply a direct correlation with the calculated $E_{chem}$ values, except that we observe a trend where Pt-Ni clusters having a small number of atoms at the cluster/oxide interface ($<$ 8) tend to display a large core strain; while those cluster with a larger number of interfacial atoms, \textit{core} strain tend to decrease up to approximately -4$\%$. Regarding \textit{shell} strain - see Fig.\ref{fig:Data} (d) - we observe that Pt-Ni clusters display smaller strain values compared to monometallic Pt clusters. There is a small window, where FCC and HCP sites on Pt$_{32}$Ni$_{6}$TO(111) as well as those for Pt$_{42}$Ni$_{13}$Ih(111) have slightly less ($<$ -1 $\%$) \textit{shell}, leading to E$_{chem}$ values lower than 0.5 eV. \\

Bader charge transfer analysis was performed for both Pt-Ni, Pt and Ni clusters at \textit{best} (strong) and \textit{worst} (weak) O$_2$ adsorption sites. In particular, Fig. \ref{fig:Data} (e) shows the amount of charge on the O$_2$ molecule calculated for the Pt-Ni, Pt and Ni systems. As a general trend, one can observe that the amount of charge transferred to the supported clusters varies as a function of interfacial metal atoms (number of contact O surface sites), with less charge being transferred from the substrate to Ni clusters, compared to Pt and Pt-Ni clusters. Overall, the O$_2$ molecule subtracts an small amount of charge ($\sim$ 0.35 e$^{-}$) per O atom at Pt-Ni and Pt sites, being larger for Ni ($\sim$ 0.5 e$^{-}$), in line with having larger E$_{chem}$ values calculated for Ni clusters. This excess of charge on the O$_2$ molecule creates a metal-superoxo (M-O$_{2}^{-}$) type intermediate (one electron transferred)\cite{TaoJPCB2001}. Charge transfer is in line with both Pt (2.28), Ni (1.91) and O(3.44) Pauling electronegativities, and to those structural motifs and interfacial geometries which maximise the number of metal atoms in contact O-surface sites. \\

There is a modest tendency of Pt-Ni clusters to reduce the overall \textit{roughness} of the (100) and (111) cluster/oxide interface layers, as shown in Fig. \ref{fig:Data} (f). In this plot, the Pt metal-metal (M-M) bond length of the interface layer in Pt-Ni clusters results to be shorter than the average M-M bond length in the cluster (Pt-Ni $\sim$ 2.65 \AA), in contrast with the pure cases (Pt $\sim$ 2.73 \AA, Ni $\sim$ 2.44 \AA), where they show an elongation on the metal distances at the interface. Even though low E$_{chem}$ values are calculated for TO and CO clusters in contact with the oxide through their large (100) facets at low \textit{roughness} (less than 0.1 $\%$), the dispersion of the overall values is too wide to provide any conclusive analysis suggesting a more local chemistry ruling O$_2$ adsorption energies. The role of the support in our work thus facilitates charge being transferred to the supported cluster (which eventually is redistributed among the cluster and O$_2$ molecule) as well as it provide an "anchor" point for the cluster. The substrates could eventually promote the controlled preparation of (111) facets as the Pt-Ni clusters grow in size, where tailored O$_2$ adsorption (and further dissociation) can take place\cite{AsaraACSCATALYSIS2016}. \\

\subsection{Inclusion of van der Waals corrections and Bader charge transfer analysis}

The inclusion of a semi-empirical dispersion correction (DFT+D) via single-point (SCF) calculations on all the inequivalent configurations involving O$_2$ adsorption have, in most cases, modestly increased the calculated E$_{chem}$ values. In general, predicted PBE energetic ordering between adsorption sites was preserved, with only a minor energetic rearrangement among the strongest chemisorption sites. Full DFT+D relaxations were also performed for all 5 chemisorption sites (Fig. \ref{fig:sites1}), only for the Pt$_{25}$, Ni$_{25}$ and Pt$_{20}$Ni$_{5}$ (TO)$_{100}$ structures to confirm the an overall preference for edge sites, such as (100)/(111) and (111)/(111) trends. For Pt-Ni clusters, DFT+D relaxations predicted the \textit{along edge} (111)/(100) as the strongest chemisorption site (1.62 eV), closely followed by \textit{along edge} (111)/(111) site (1.47 eV) just as the PBE energetic ordering. Similar gains in energy are calculated for the FCC and HCP sites. For the Pt clusters, the best adsorption site is now the \textit{along edge} (111)/(100) instead of the predicted PBE \textit{along edge}(111)/(111), at 1.96 eV. Being now the O$_2$ molecule at close proximity to the oxide surface (2.99 \AA), DFT+D is able to capture an stronger overlap between the electron orbitals of O$_2$ and those of the oxide surface. This results in an E$_{chem}$ increase up to 1.96 eV, from the 1.38 eV single-point (DFT+D) calculation, an increased also calculated for the \textit{between edge}(111)/(100)site, from 1.67 to 1.71 eV. Analogous trends are observed for Ni clusters, where \textit{along edge} (111)/(100) site remains as the one having the largest E$_{chem}$ value (2.55 eV), while marginal gains are calculated for sites \textit{along edge} (111)/(111) and \textit{between edge} (111)/(100) (2.25 and 1.79 eV, respectively). \\

\subsection{Electronic vs. geometric properties}

Upon O$_2$ adsorption, we calculated the projected density of states (PDOS) for the \textit{best} (strongest) and \textit{worst} (weakest) - namely the FCC site - adsorption sites, see Fig. \ref{fig:dcenters}. Included within the figure, is the calculated \textit{d}-band center\cite{VojvodicDBAND2014}. This is done for those Pt atoms (and Ni) involved in the O$_2$ adsorption (which can occur, both molecularly and dissociated). Comparisons between different Pt-Ni clusters are difficult due to the size range considered as well as some \textit{worst} configurations involve dissociated O$_2$. Overall, we observe that the calculated \textit{d}-band centers of the 2 Pt atoms from the \textit{best} adsorption sites (molecularly adsorbed O$_2$) are positioned nearly at the same energy values (see vertical black lines, Fig.\ref{fig:dcenters}). For the all the \textit{best} sites, this implies having their corresponding \textit{d}-centers positioned lower in energy with respect to the Fermi energy ($E_{f}$), thus accounting for stronger interactions with the molecule O atoms, except for those configurations where O$_2$ is dissociated, namely Pt$_{20}$Ni$_{5}$TO(100), Pt$_{32}$Ni$_{6}$TO(100) and Pt$_{42}$Ni$_{13}$CO(100), FCC sites). For most of these dissociated configurations, we observe that the position of the Pt \textit{d}-centers, of the least coordinated Pt atom is highly shifted to low energies thus implying a strong interaction with the O atom; see Pt$_{32}$Ni$_{6}$TO(100) and Pt$_{42}$Ni$_{13}$CO(100). Most the rest of the \textit{worst} configurations, the Pt \textit{d}-centers are closer to the $E_{f}$, thus corresponding to weaker interactions between the O$_2$ and the cluster. However, Pt$_{32}$Ni$_{6}$TO(111) is an exception, as the O$_2$ is adsorbed in a molecular fashion, though the the central Pt \textit{d}-centers - occupying the (111) facet where adsorption takes - place is shifted to lower energy values as the Pt atom is "lifted" from its position as it interacts with the O$_2$ molecule, a geometric effect highlighted in previous work\cite{JenningsNANOSCALE2014}. Larger clusters, such as Pt$_{42}$Ni$_{13}$ (both CO and Ih) and the decahedral Pt$_{47}$Ni$_{11}$, further follow similar \textit{d}-centers shift trends. However, close overlaps between calculated \textit{d}-centers for \textit{worst} and \textit{best} sites, particularly for Pt$_{47}$Ni$_{11}$, are observed as $E_{chem}$ differences for the two sites become smaller. 

Thus, to get disentangle electronic from geometric effects, we have calculated the corresponding GCN for both adsorption sites for all Pt-Ni clusters as done for the \textit{d}-band centers. The calculated GCN values, as described in Eq. \ref{eq3}, show for Pt-Ni clusters an overall linear relationship with respect to calculated $E_{chem}$ values; see Fig. \ref{fig:GCN}. As the GCN value increases, a reduction on the $E_{chem}$ values is seen, particularly above a GCN $>$ 8, where most of the calculated values involving FCC and HCP sites resemble the experimental and those calculated values for O$_2$ adsorption on Pt(111) surface (0.46 up to 0.86 eV)\cite{GlandSURFSCIE1980, SteiningerSURFSCIE1982A, SautetPRB1999, EichlerPRL1997, JenningsNANOSCALE2014, McEwenPCCP2012}, and even below. There was only 1 case where the FCC site reported a lower GCN value (6.722), on the the Pt$_{32}$Ni$_{6}$(111) cluster. Here, the central Pt atom of the (111) facet was lifted with respect to the facert thus separating it from the rest of the core Pt atoms and thus lowering its GCN value. On the opposite, the strongest $E_{chem}$ values ($>$ 0.64 eV and above) can be found at sites such as \textit{along edges} (111)/(111); where GCN between 4 and 5 are calculated. For monometallic Pt clusters, a similar trend is observed. However, the slope of the linear fitting is less pronounced due to the "soft" nature of the Pt-Pt bonds within the cluster, which can stretch their cluster M-M bonds to accommodate the incoming O$_2$ molecule effecting the overall GCN and $E_{chem}$ values. Ni clusters on the other hand, interact their strongly with O$_2$ that for GCN values between 5 and 6 ($\sim 2 eV$) - leaving a narrow window from 6 to 8 for a slightly weaker adsorption - again at GCN above 8, the O$_2$ molecule is strongly adsorbed and in some cases fastly dissociated thus affecting the overall linear trend. Interestingly, we have reported similar trends where O$_2$ adsorption can take place at low $E_{chem}$ ($<$ 0.5 eV) for GCN values above 8, for the particular case of larger TO and CO clusters involving 82 and 86 atoms, respectively\cite{AsaraACSCATALYSIS2016}. 

\section{Conclusions}

Using a wide variety of Pt-Ni, Pt and Ni geometries, sizes and compositions supported over MgO(100) our results we shown that molecular have shown that O$_2$ adsorption on the sub-nanometer range (from 25 up to 58 atoms) is a rather complex case. Their corresponding electronic structure as analysed via the calculated position of \textit{d}-band center as well as the local geometry at the adsorption site using the generalised coordination number (GCN). The calculation of these two quantities show that these two effects are deeply intertwined, directly affecting the calculated E$_{chem}$ values. O$_2$ adsorption is regulated by mainly 2 types of inequivalent sites: (a) strongest E$_{chem}$ values (above 1 eV) are calculated at cluster edges, such as the \textit{along edge}(111)/(111) and (111)/(100) as well as the \textit{between edge}(111)/(100) sites which tend to have a low GCN value ($< $6); and (b) the FCC and HCP sites where the \textit{weakest} E$_{chem}$ values (less than 0.64 eV) are calculated, involving GCN values $>$ 8 which resemble those O$_2$ adsorption energies of the flat Pt(111) surface. In general, calculated \textit{d}-band centers positions for those Pt atoms at the adsorption are located at lower energies from the Fermi level (E$_f$), thus indicating a stronger binding between the metal and the O$_2$ adsorbate. The opposite is seen for the weakest adsorption sites, where \textit{d}-band centers positions are closer to the E$_f$; though this trend can be affected if the O$_2$ is found to be dissociated as Pt interactions with atomic O are stronger compared to the molecular case.

From our calculations provide - as far as we are concerned - the first O$_2$ chemisorption map for across a wide range of supported Pt-Ni clusters. Clear O$_2$ adsorption trends provide a deeper understanding on potential adsorption sites for low-barrier O$_2$ dissociation, following Sabatier principle. This will imply that the best catalysts should be the ones that bind both atoms and molecules not too weakly nor too strongly, in order to activate the reactants and to efficiently desorb the products\cite{NoskovCHEMREV2008}. We show from our results that it is theoretically possible to construct core-shell systems, where the core of the nanoparticle (cluster) is made of a cheaper metal (Ni) compared to a more expensive metal on the shell (Pt). This not only reduces the amount of Pt-loading by alloying; but preserves the Pt chemistry at a mere 1 atomic-layer Pt-shell where O$_2$ adsorption process can occur with calculated E$_{chem}$ values of the order of those seen for a full monometallic Pt nanoparticle and bulk Pt(111) surface. Having systematically analysed the effects of O$_2$ adsorption over supported over mono (Pt, Ni) and bimetallic (Pt-Ni) clusters ($\textless$1nm.), we foresee that a potential tailoring of chemisorption properties and the rational design of novel nanocatalysts will ultimately depend a synergistic combination of controlled NPs preparation methods (size, geometric and segregation effects) and on a profound knowledge of the complex electronic properties of both NPs and oxide support.

\section{Acknowledgements}

This work has been supported by the UK research council EPSRC under the Critical Mass TOUCAN project, grant number EP/J010812/1. FB benefit of the financial support by EPSRC, under Grant No. EP/GO03146/1 and the Royal Society grant No. RG120207. Density Functional Theory (DFT) calculations have been performed on ARCHER (Cray XC30 system), the UK's National High-Performance Computing (HPC) Facility. LOPB acknowledges Superc\'{o}mputo UNAM (Miztli) for computational resources under projects SC16-1-IG-78 and SC15-1-IG-82 as well as financial support from PAPIIT-UNAM (Project IA102716). 

\footnotesize { 
\bibliography{ptni_mgo} 

\bibliographystyle{rsc} 

}

\end{document}